\documentclass[%
 aip,
 jcp,
 amsmath,amssymb,
 reprint,%
]{revtex4-1}

\usepackage{graphicx}
\usepackage{dcolumn}
\usepackage{bm}

\usepackage[utf8]{inputenc}
\usepackage[T1]{fontenc}
\usepackage{mathptmx}

\begin{document}


\title{Assessing the Accuracy of the Jastrow Antisymmetrized  
Geminal Power in the $H_4$ Model System}

\author{C. Genovese}
 \affiliation{SISSA -- International School for Advanced Studies, Via Bonomea 265, 34136 Trieste, Italy;}
 \email{claudio.genovese@sissa.it}
\author{A. Meninno}%
\affiliation{Dipartimento di Fisica -- Universit\`a di Pisa, Largo Bruno Pontecorvo 3, 56126 Pisa, Italy;}
\author{S. Sorella}
\affiliation{SISSA -- International School for Advanced Studies, Via Bonomea 265, 34136 Trieste, Italy;}
\email{sorella@sissa.it}
\affiliation{Democritos Simulation Center CNR--IOM Istituto Officina dei Materiali, 34151 Trieste, Italy.}


\date{\today}

\begin{abstract}
We report a quantum Monte Carlo (QMC) study,  on a very simple but nevertheless very instructive model system of four 
hydrogen atoms, recently proposed in Ref.~\onlinecite{doi:10.1063/1.4986216}. We find that the 
Jastrow correlated Antisymmetrized Geminal Power (JAGP) is able to recover most of the correlation energy 
even when the geometry is symmetric and the hydrogens lie on the edges of a perfect square. Under such 
conditions the diradical character of the molecule ground state prevents a single determinant ansatz 
to achieve an acceptable accuracy, whereas the JAGP  performs very well for all geometries. 
Remarkably, this is obtained with a  similar computational effort. Moreover
we find  that the Jastrow Factor is fundamental in promoting  the correct 
resonances among several configurations  in the JAGP, that  cannot show up 
in  the pure 
Antisymmetrized Geminal Power (AGP). We also show the extremely fast convergence of this approach
in the extension of the basis set.  
Remarkably only the simultaneous optimization of the Jastrow and the AGP part of 
our variational ansatz 
is able to recover an almost perfect nodal surface, yielding therefore 
 state of the art energies, almost converged in the complete
 basis set limit (CBS), when the so called Diffusion Monte Carlo is applied. 
\end{abstract}

\maketitle

In recent years much progress has been made in the definition of variational wave functions (WF)  
capable to describe rather accurately the electron correlation. To this purpose two strategies have been employed:
i) the use of multi-determinant wave functions 
\cite{doi:10.1063/1.471865,0034-4885-74-2-026502,doi:10.1021/ct3003404} 
or ii) exploiting the large variational 
freedom that can be achieved by applying a correlation term, dubbed Jastrow factor (JF), to a generic pairing function
\cite{doi:10.1063/1.1604379,PhysRevB.77.115112,PhysRevLett.96.130201}. Even if the latter approach  cannot be 
systematically improved, it may open the way to deal
with large systems, thanks to the moderate scaling with the number of electrons. Indeed, the corresponding 
correlated WF, can be simulated efficiently within a statistical method, based on quantum Monte Carlo
\cite{RevModPhys.73.33}. Thanks to well established advances 
\cite{sr_me,linear} in this field, it is possible nowadays 
to compute  the total  energy of a given correlated ansatz and to optimize several variational parameters 
with a computational effort scaling at most with the fourth power of the number of electrons.    

A good variational ansatz allows a good description of the ground state by energy optimization. Moreover an
even better characterization can be obtain by applying the so called diffusion Monte Carlo (DMC) method with the
Fixed Node approximation (FNA)\cite{doi:10.1063/1.4954726, lester1997recent}. 
Within this projection method it is possible 
to obtain the lowest energy state constrained to have 
the same signs of a chosen  trial WF, in the configuration space where electron 
positions and spins are given.
The connected regions of space with the same sign are called 
nodal pockets and the surface determining this pockets, satisfying WF$=0$, the nodal surface.
Usually the energy optimization, implemented here, has been shown to be very 
successful to determine the nodal surface of the WF as we will show also in the present study. In this 
work  we have used a particular method for implementing the FNA  that is called lattice regularized diffusion Monte 
Carlo (LRDMC)\cite{doi:10.1063/1.3516208}. While usually a short imaginary time approximation is applied to the 
propagator for its finite time evaluation, in the LRDMC a lattice regularization is employed in the physical space
by using a finite mesh approximation\cite{doi:10.1063/1.3380831}.

A widely used ansatz for the WF is the single Slater determinant (SD).
The SD can be taken directly from Hartree-Fock (HF) or density functional theory (DFT) 
calculations without further optimizations. Unfortunately this approach can fail to describe
the exact ground state WF and its nodes and therefore a more complex procedure is often required. Depending on the problem it may be necessary
 to further optimize the SD parameters or to change the ansatz. Instead of the SD
it is possible to consider more advanced WF using multideterminant ansatzes within, for instance, Full Configuration Interaction (FCI) or  
 Complete Active Space CAS($n,m$), with $n$ active electrons in $m$ orbitals, 
or a more accurate single determinant  Geminal WF (AGP)\cite{doi:10.1063/1.1604379, doi:10.1063/1.1794632, PhysRevLett.109.203001, doi:10.1063/1.4829536, doi:10.1063/1.4829835, PhysRevB.84.245117}. It is well known that it is possible to improve considerably the
 correlation energy by multiplying a given ansatz by a JF. The WF built with
an AGP and a JF is indicated as JAGP, while the SD and the JF as JSD. Once 
the ansatz of the WF is given it is also important to choose an appropriate atomic basis
set. Enlarging the basis set allows us to be closer to the CBS limit, 
but at the same time increases the number of variational parameters.  It is therefore 
important to have a compact description of the WF by using an atomic basis set as small as 
 possible.

Even relatively simple systems can hide pitfalls that can be very difficult to solve. The case of the 
(H$_2$)$_2$, a system of two diatomic molecule of hydrogen at equilibrium distance, first introduced in
QMC literature by Anderson \cite{Anderson:1975,doi:10.1002/qua.560150111}, is emblematic from this point of view: as recently
 shown by Gasperich et al. \cite{doi:10.1063/1.4986216}, a single SD can only give a very poor description 
of this system when it  approaches the square geometry. This is due to the degeneracy of the frontier orbitals 
in the square limit that a single SD is not able to reproduce. In this 
paper we show that a single AGP determinant enriched with the JF correlator allows a perfect 
description of the ground state. 

The simplicity of this model system allows us to study the role of the optimization 
in determining an accurate nodal surface, because, by repeating several times the 
optimization, we can be safely confident that the absolute minimum energy WF is obtained.
On the other hand we can also verify that our stochastic optimization\cite{becca2017quantum} 
works also when we remove the Jastrow from our ansatz,  providing  the lowest energy 
AGP, clearly with much larger computational effort compared to deterministic 
methods, that,  to our knowledge, are not available for the AGP.
This tool has been proved to be very useful in this work because we are able to show that
the use of  a pure  AGP determinant (without any JF)  can give rise 
not only to a poor description of the electronic correlation, but also a qualitatively 
wrong picture of the chemical bond.

Moreover, similarly to what found 
in the benchmark study of the hydrogen chain \cite{PhysRevX.7.031059}),
the fullfillment of the electron-electron and electron-ion cusp conditions, obtained with a 
suitably chosen JF, 
makes the convergence to the CBS particularly fast and efficient
, requiring only 
a double zeta gaussian basis set (cc-pVDZ) for the accurate description of the corresponding nodal surface.  
Remarkably,  the DMC energies obtained with the double zeta JAGP trial WF  
are better than the ones obtained with the CAS(2,2) and CAS(4,4) and also with
the full configuration interaction (FCI) calculated with a quadruple zeta basis \cite{doi:10.1063/1.4986216}.

\section{Wave functions and procedure} \label{sec.methods}

For this study we used WFs given by the product of a determinant, SD or AGP, and a JF, optimized
with standard stochastic techniques\cite{doi:10.1063/1.3516208}. 
For all the calculations we used the TurboRVB
package.\cite{webpage} 

The value of the WF for an electronic configuration $\mathbf{X}$ is given as
\begin{equation}
  \langle \mathbf{x} | \Psi \rangle = \Psi(\mathbf{X})= J(\mathbf{X})\times \Psi_{det}(\mathbf{X})
  \label{WF}
\end{equation}
We will firstly describe the determinant part of the WF, moving then to the description of the JF
used. 

In TurboRVB we use an atom-centered basis set of $N_{orb}$ orbitals $\left \{ \phi_{I, \nu}(\mathbf{r})  \right \}$, where
$I$ and $\nu$ indicate the $\nu$-th orbital centered on the $I$-th atom at the  position 
$\mathbf{R}_I$. For this study standard 
gaussian basis sets are  
chosen for the determinant, with orbital types: 
\begin{equation}
  \phi_{I,\nu}(\mathbf{r})=e^{-\frac{|\mathbf{r}-\mathbf{R_I}|^2} {Z_\nu}} Y_{l,m}^\nu,
  \label{basis}
\end{equation}
where $Z_\nu$ is a numerical coefficient that describes how diffuse the atomic orbital is around the atom, while $Y_{l,m}^\nu$
is the spherical harmonic function relative to the orbital type of $\nu$. For the sake of compactness one can 
enumerate the basis as $\left \{ \phi_{\mu}(\mathbf{r})  \right \}$ combining the indices $\nu$ and $I$ in 
a single index $\mu$ for a lighter notation. The use of a double zeta basis indicates that for the description of the $1s$ 
orbital of the hydrogen we are using the $s$-waves and the $p$-waves orbitals, while for the triple zeta we are using 
also  $d$-waves orbitals.

In order to introduce the AGP WF we first define a singlet wavefunction of an electron pair:
\begin{equation}
  \psi_2( \mathbf{r}_1 \sigma_1,\mathbf{r}_2 \sigma_2) = { 1 \over \sqrt{2}} ( |\uparrow \downarrow\rangle 
  - |\downarrow \uparrow \rangle) f( \mathbf{r}_1,\mathbf{r}_2)
  \label{singlet} 
\end{equation}
where $f$ is the so called geminal function, that is assumed to be symmetric, namely
$f( \mathbf{r}_1,\mathbf{r}_2)=f( \mathbf{r}_2,\mathbf{r}_1)$, for a perfect singlet.
The geminal function $f$ is expanded as 
\begin{equation}
  f(\mathbf{r}_1,\mathbf{r}_2)=\sum_{k,l} \lambda_{k,l} \phi_k(\mathbf{r_1})
  \phi_l(\mathbf{r_2}).
  \label{geminal}
\end{equation}

The generalization to a many electron WF requires an antisymmetrization of $N/2$ electron
pairs of the form given in Eq.~(\ref{singlet}). The value of the WF for a given electronic configuration 
is given by the determinant of the matrix $F$
 \begin{equation}
   \langle x | AGP \rangle =  \det F
   \label{agpdet}
 \end{equation}
where the $N/2 \times N/2$ matrix $F$ is defined as $F_{i,j}=f(\mathbf{r}_i,\mathbf{r}_j)$, with the row index $i$
corresponding to the up electrons and the column index $j$ to the down ones.
In the case in which $N_{\uparrow}\ne N_{\downarrow}$, we assume, without loss of generality that $N_{\uparrow}> N_{\downarrow}$.
If it is then necessary 
we add $N_{\uparrow}-N_{\downarrow}$ columns to the matrix, corresponding to the unpaired orbitals written in the same  basis set.  
Thus the matrix is a $N_\uparrow \times N_\uparrow$ square matrix and its determinant can be evaluated, yielding also in this case the value of the AGP.

For our calculations we often  initialize the AGP or the SD WFs starting from  
a DFT calculation \cite{becca2017quantum}, within the LDA approximation. Thus it is important to translate without loss of 
information the SD into an AGP. 

A SD is characterized by its set of MOs
\begin{equation}
  \Phi^{mol}_\alpha(\mathbf{r})=\sum_{i=1}^{N_{orb}}P_{\mu,\alpha}\phi_\mu(\mathbf{r}),
  \label{molorb}
\end{equation}
that uniquely define it.

We can recast the SD into an AGP whose matrix $F$ is diagonal in the basis of the MOs. The obtained geminal function 
can be written as
\begin{equation}
  f(\mathbf{r},\mathbf{r}')=\sum_\alpha \bar{\lambda}_{\alpha,\alpha} \Phi^{mol}_\alpha(\mathbf{r}) \Phi^{mol}_\alpha(\mathbf{r}'),
  \label{initwf}
\end{equation}
where the diagonal form of the matrix $\bar{\lambda}$ is ensured by the orthogonality of the MOs. If we substitute the 
expression (\ref{molorb}) in (\ref{initwf}) we can have the description of the pairing function in the originally chosen
basis set
\begin{equation}
  \lambda=P_{\uparrow}\bar{\lambda}P_{\downarrow}^T.
  \label{initlambda}
\end{equation}
In this way the SD is translated into an AGP of the same form given in  Eq.~(\ref{agpdet}),  where the geminal functions
 have the same expression of Eq.~(\ref{geminal}) once the matrix (\ref{initlambda}) is substituted into Eq.~(\ref{initwf}). 
The SD written in this way appears as  an AGP but for each configuration $\mathbf{x}$ the WF value is unchanged. 

When we initialize the WF from a set of MOs, we have the same value for the AGP and the SD and the differences between
the ansatzes can be seen only after the energy optimization. The AGP always provides an 
energy lower than the SD one, due to the larger variational freedom. 
Indeed, in the case of the SD, 
the matrix $\lambda$, is constrained to have  
only  $N/2$  orthogonal MOs\cite{doi:10.1063/1.3249966}, corresponding therefore to a number of  variational 
parameters much smaller than the one required for the AGP.
\begin{figure}
	\centering
	\includegraphics[scale=0.25]{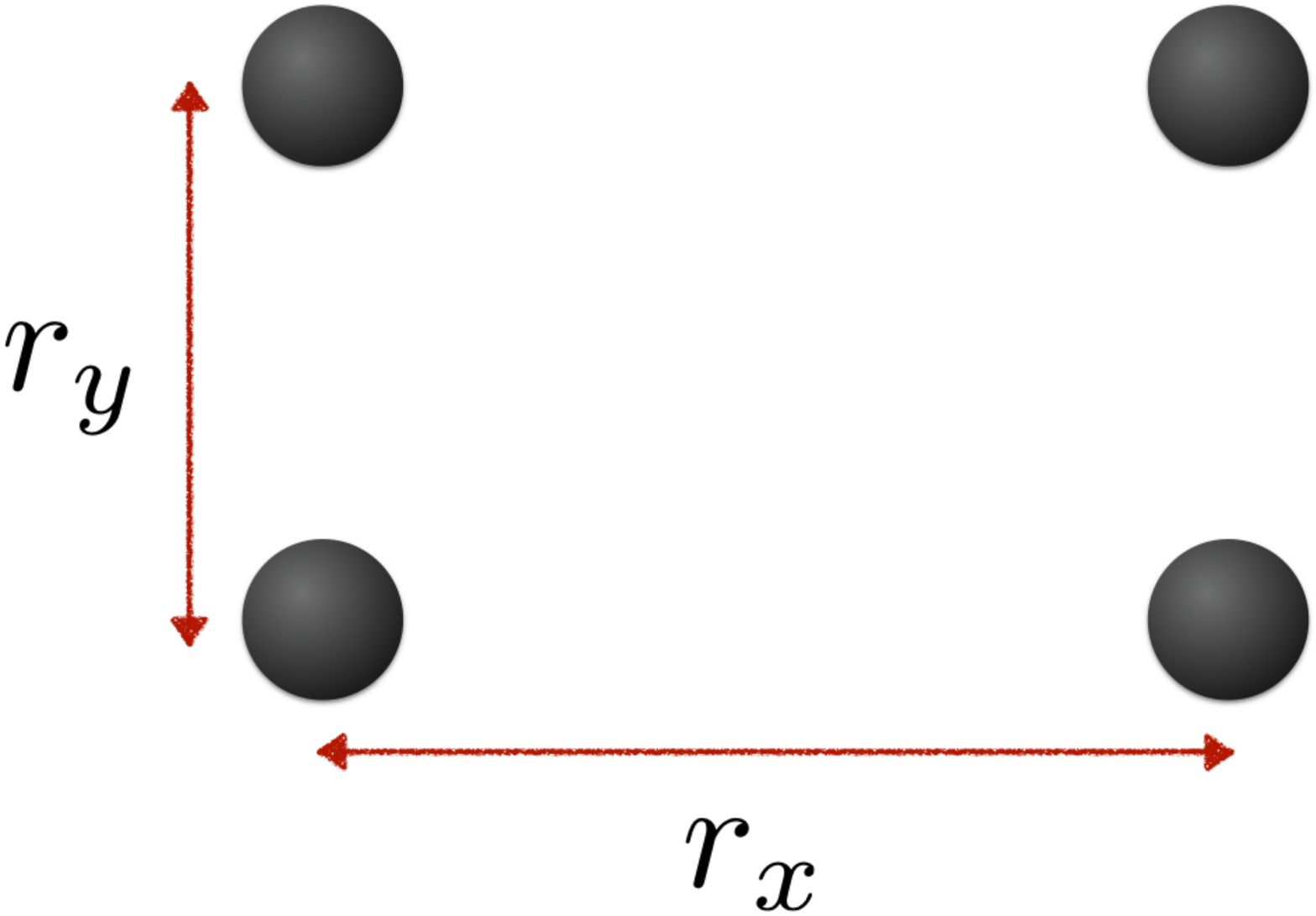}
	\caption{Stylized picture of the system. While $r_y$ is kept constant for all the calculations at a distance equal to 
        $2.4$ a.u., the distance $r_x$ is varied for different system shapes. }
	\label{system}
\end{figure}

The two different WFs that we have introduced do not describe well the correlation 
between the electrons. Within QMC, it is easy  to improve the quality of the WF, by  multiplying the SD and the AGP
 with an exponential JF. This factor has also another important effect,
as it speeds up the  convergence to the CBS by improving  the description of the atomic core, and thus not requiring
 large $Z_\nu$ values in the basis set. In particular our JF is in the form
\begin{equation}
  J(\mathbf{X})=e^{U_{ei}+U_{ee}},
  \label{JF}
\end{equation}
where $U_{ei}$ is a single body term dealing explicitly with the electron-ion interaction and  $U_{ee}$ is a many-body term
to take into account the electronic correlation. 
 The JF is particularly useful because, with an appropriate choice,
 it is possible  
to satisfy exactly  the 
 electron-electron and electron-ion cusp conditions of the many-body WF, a consequence  
of the Coulomb $1/r$ singularity at short distance.
The single body term is indeed written in the form 
\begin{equation}
  U_{ei}=\sum_{i=1}^{\#el}u_{ei}(\mathbf{r}_i),
  \label{sb1}
\end{equation}
with $u_{ei}$ being
\begin{equation}
  u_{ei}(\mathbf{r}_i)=- \sum_{I=1}^{\#ions}Z_I \frac{1-\exp(b_{ei}|\mathbf{r}_i-\mathbf{R}_I|)}{b_{ei}},
  \label{sb2}
\end{equation}
where $Z_I$ is the atomic number of the atom $I$ and $b_{ei}$ is a variational parameter.
The electron-electron term is written as
\begin{equation}
  U_{ee}=\sum_{i<j}u_{ee}(\mathbf{r}_i,\mathbf{r}_j),
  \label{mb1}
\end{equation}
where the sum is extended over the pairs of different electrons and where
\begin{equation}
  u_{ee}(\mathbf{r}_i,\mathbf{r}_j)=\frac{|\mathbf{r}_i-\mathbf{r}_j|}{2(1+b_{ee}|\mathbf{r}_i-\mathbf{r}_j|)}
+\sum_{k,l}g_{k,l}\bar{\phi}_k(\mathbf{r}_i)\bar{\phi}_l(\mathbf{r}_j).
  \label{mb2}
\end{equation}
The first term in the Eq.~(\ref{mb2}) deals explicitly with the cusp conditions of the electron-electron potential with
$b_{ee}$ as variational parameter, the second term instead takes explicitly into account the correlation via a pairing
function in the same form of Eq.~(\ref{geminal}) with the matrix $g$ as variational parameters. Here the sum over $k$ and
$l$ is extended over a  basis set similar  to the one 
used for the determinant, namely determined by 
 atomic-like wave functions of the form
\begin{equation}
  \bar{\phi}_{J,\mu}(\mathbf{r})=e^{-\frac{|\mathbf{r}-\mathbf{R_J}|} {Z_\mu}} Y_{l,m}^\mu,
  \label{basis}
\end{equation}
with $\mu$ denoting the orbital  type and $R_J$  the nuclear positions of the atoms considered; as already done in the 
Eq.~(\ref{mb2}) we combine $\mu$ and $J$ in a single index.

The geometry of the system studied has a fixed bond distance along the $y$ direction $r_y=2.4$ a.u.. This value gives the lowest energy 
result for the square geometry \cite{doi:10.1063/1.1680481}. As sketched in Fig.~(\ref{system}), we study the system  as a function of  the distance $r_x$ between the two vertical molecules.  

For the optimization of the JAGP and JSD WFs we used the same procedure. We consider two types of initializations that 
we denote in the following by OPT $r_x > r_y$ or OPT$r_x < r_y$, to indicate that the tetragonal symmetry is broken. In the first
 (second)  case we take $r_x=4$a.u. ($r_x=1.8$a.u.) and perform  a DFT calculation  for the initial Slater determinant while the 
Jastrow is all zero but  the initial one and two body parameters are  set to a non vanishing value $b_{ee}=1/2$ and $b_{ei}=1.3$. 
 We initially optimize only the JF and we proceed
with the full optimization of the AGP or SD with the JF. Then we move the atoms to a new position close to the original one
mantaining the values of the variational parameters. If the new solution is reasonably close to the previous one, the stochastic
optimization drives the WF to its new minimum. We iterated this procedure to obtain the WFs at all the $r_x$ distances for the
JAGP and JSD. As we will discuss more extensively later, the JAGP optimization does not depend much on the starting WF, 
that  is instead crucial  for the JSD. In this latter case the optimization procedure determines 
completely different results depending on the initial geometry, when we get close to the symmetric square case.

For the optimization of the AGP without the JF we followed two different procedures yielding  the same results. In one case
we started for every geometry from the corresponding optimized JAGP WF: we set the JF to 1 and we optimized the AGP from there. 
In the second case we have used the same  procedure adopted for the 
the JAGP and JSD cases and obtained consistent energy values, validating the optimization 
procedure even in this difficult case without the Jastrow factor. 
\begin{figure}
	\centering
	\includegraphics[scale=0.8]{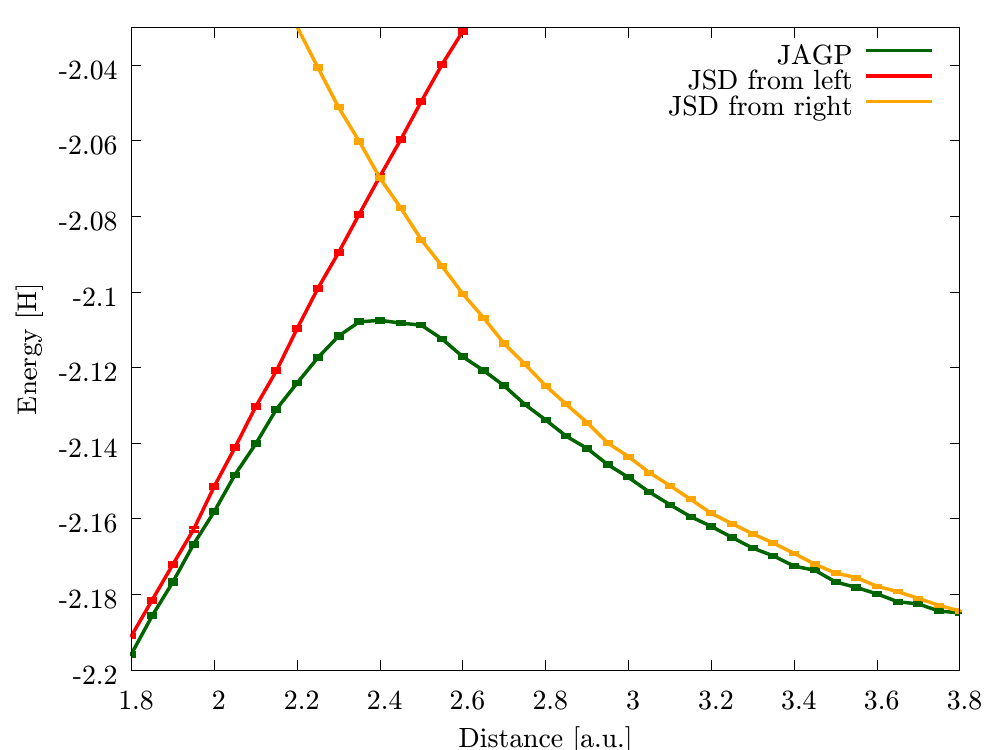}
	\caption{Comparison between the variational energies of the different WFs. In orange the energies of the JSD starting from the
        calculation at large $r_x$, in red the ones  starting from the small $r_x$, while in green 
        the JAGP variational energies are reported.}
	\label{varen}
\end{figure}
\begin{table}
  \caption{Variational energies for different optimized WFs. The basis set used for the AGP and SD is indicated between round 
    parenthesis. We show one point for each case: $r_x=r_y$, $r_x < r_y$ and $r_x > r_y$. All the energies are expressed in Hartree.}\label{tab:varen}

  \begin{tabular}{l c c c }
    \hline
    \hline
    \rule[-0.4mm]{0mm}{0.4cm}
    $r_x$ & JSD(cc-pVDZ) & JAGP(cc-pVDZ) & JAGP(cc-pVTZ) \\
    \hline
    \rule[-0.4mm]{0mm}{0.4cm}
    $1.80$& $-2.1909 \pm 0.0003$ & $-2.1957 \pm 0.0004$ & $-2.1953 \pm 0.0003$ \\
    \rule[-0.4mm]{0mm}{0.4cm}
    $2.40$& $-2.0694 \pm 0.0004$ & $-2.1075 \pm 0.0004$ & $-2.1084 \pm 0.0003$\\
    \rule[-0.4mm]{0mm}{0.4cm}
    $3.00$& $-2.1435 \pm 0.0003$ & $-2.1491 \pm 0.0003$ & $-2.1504 \pm 0.0003$ \\
    \hline
    \hline
\end{tabular}			
\end{table}
\begin{figure}
    \includegraphics[scale=0.95]{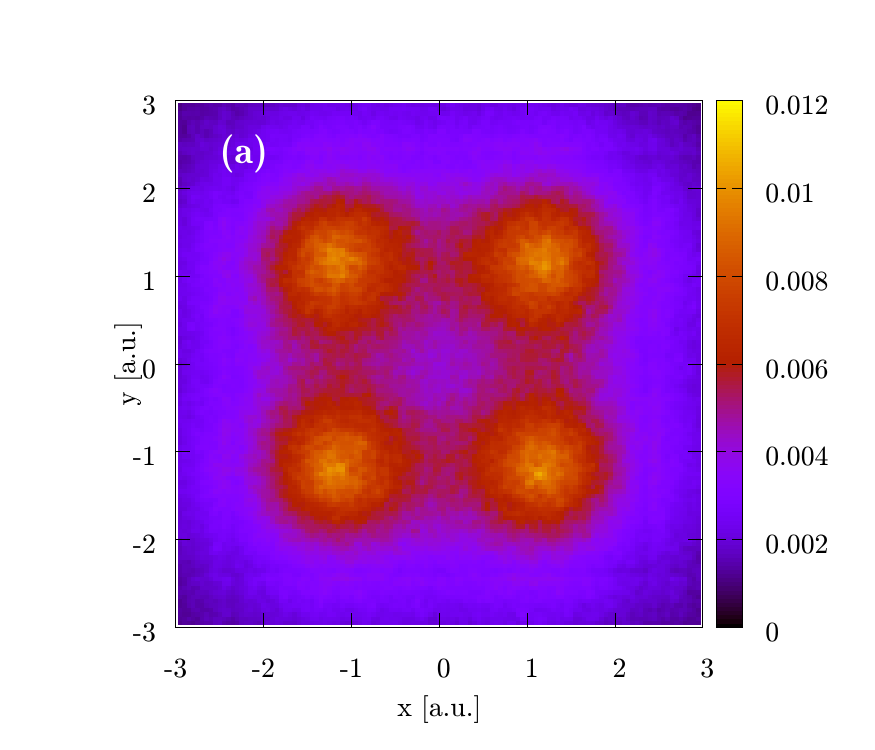}
    \includegraphics[scale=0.95]{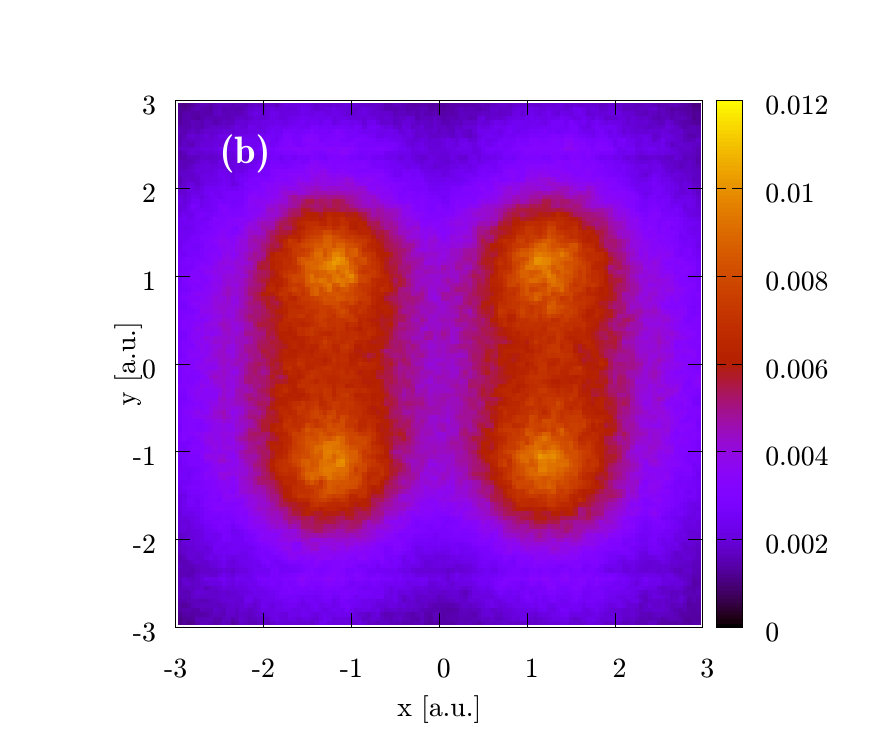}
    \includegraphics[scale=0.95]{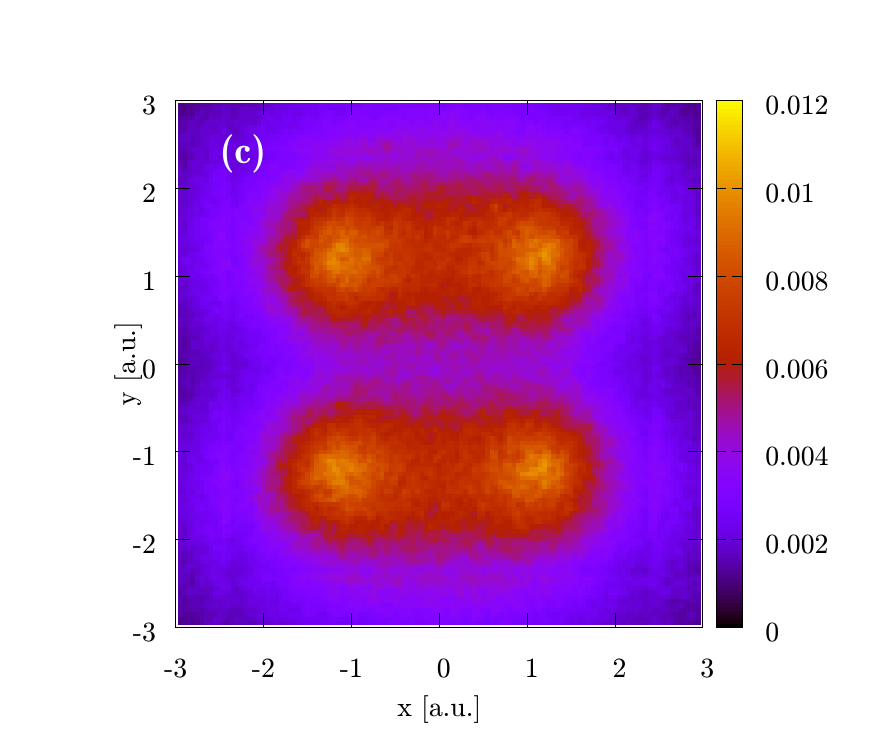}
  \caption{Density plot on the $xy$ plane of the systems for the square geometry for the JAGP and JS WFs. 
    In panel (\textbf{a}) the density obtained with the JAGP WF, in the panel (\textbf{b}) the density of the JSD optimized from
    $r_x>r_y$, while in the panel (\textbf{c}) the one of the JSD from $r_x<r_y$.}
  \label{denspl}
\end{figure}
\begin{figure}
	\centering
	\includegraphics[scale=0.8]{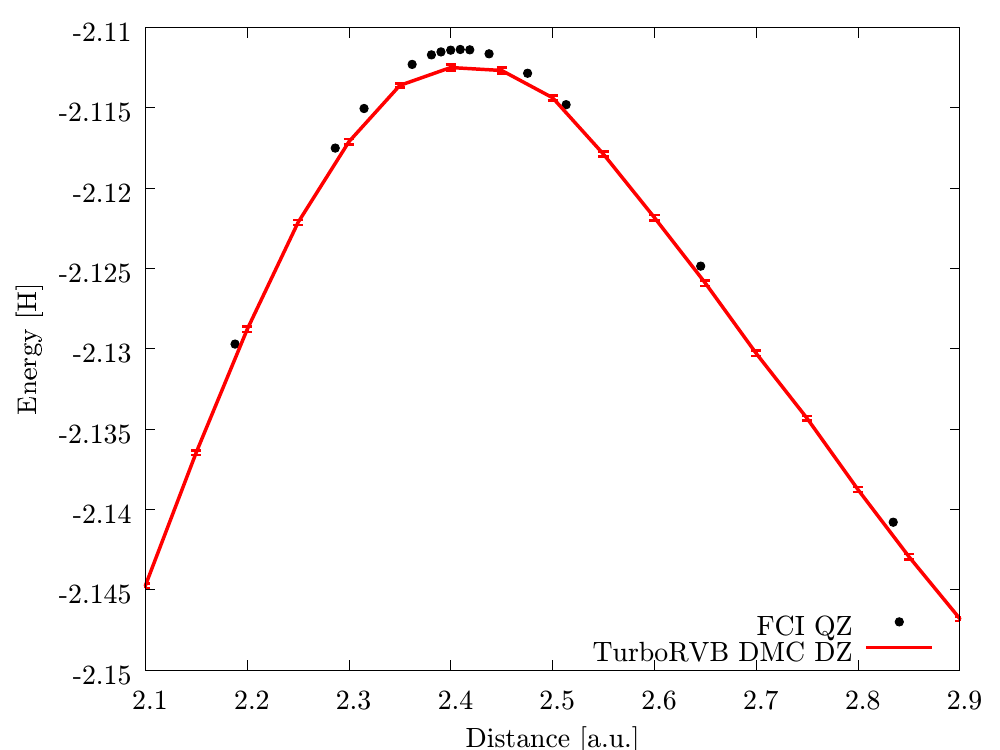}
	\caption{Comparison between the DMC energy calculated using the node of the JAGP (cc-pVDZ) and the FCI (cc-pVQZ) calculation. On
        this scale the error bars of the DMC calculation are not visible.}
	\label{fnen}
\end{figure}
\begin{table}
  \caption{Difference between the energies calculated with the DMC performed using the nodes of the JAGP, the ones 
    of the CAS(4,4) and the FCI \cite{doi:10.1063/1.4986216}. All the energies are expressed in Hartree.}\label{tab:fnen}
  \begin{tabular}{l c c c }
    \hline
    \hline
    \rule[-0.4mm]{0mm}{0.4cm}
    $r_x$ & JAGP & CAS(4,4)  & FCI \\
    \hline
    \rule[-0.4mm]{0mm}{0.4cm}
    $2.188$& $-2.1307 \pm 0.0001$ & $-2.13033 \pm 0.00010$ & $-2.1297$ \\
    \rule[-0.4mm]{0mm}{0.4cm}
    $2.400$& $-2.1125 \pm 0.0002$ & $-2.11193 \pm 0.00005$ & $-2.1114$ \\
    \rule[-0.4mm]{0mm}{0.4cm}
    $2.646$& $-2.1257  \pm 0.0001$ & $-2.12558 \pm 0.00003$ & $-2.1248$ \\
    \hline
    \hline
  \end{tabular}		
\end{table}

\section{Results and Discussion}

The variational energies for the considered WFs are visible in Fig.~(\ref{varen}) and reported 
in Table (\ref{tab:varen}). 
As shown in  Fig.~(\ref{varen}), the JSD values are reasonably accurate when the system is far from 
the square geometry, but very poor when $r_x \approx r_y$. 
 We notice that for the JSD the starting point is fundamental and the optimization result can significantly differ 
depending on the two different initializations. 
A particularly evident effect is 
the crossing of the JSD energy dispersions 
in Fig.~(\ref{varen}). As expected this problem  does not affect the JAGP WF that shows the correct  profile because 
, close to the square geometry contains implicitly the two important Slater determinants with strong bonds either in the 
$x$ or in the $y$ direction. The optimizations
of the JAGP both from OPT$r_x > r_y$ and OPT$r_x < r_y$ lead exactly to the same result. The qualitative difference between the
two ansatzes is clearly shown in  Fig.~\ref{denspl}. The MOs try to localize the charge between two pairs of atoms to form 
two H$_2$ molecules. 
In particular the JSD binds the atoms that are at smaller distances in the initial geometry: if we consider the  OPT$r_x>r_y$ case 
we obtain
an higher charge density along the $y$ direction, while if we start from the OPT$r_x<r_y$ case 
 an higher charge along the $x$ direction shows up. The JAGP, instead,
can resonate between these two configurations and catch the resonance valence bond 
(RVB)\cite{rvb_pauling}  behaviour
expected for the ground state of the square geometry.

The JAGP result is not only good at the variational level, but it provides also particularly accurate nodal surfaces for the DMC calculations. 
Indeed, as we can notice from Fig.~(\ref{fnen}) and from Table (\ref{tab:fnen}), the DMC energies calculated using the nodes of 
the JAGP (cc-pVDZ) are lower then the ones calculated with the multi-determinant WF CAS(4,4), 
and FCI with the quadruple zeta basis \cite{doi:10.1063/1.4986216}. This shows that even with a small basis set the JAGP leads,  
in this controlled case, to almost  optimal nodes and, by consequence, very accurate DMC energies. This is indeed remarkable, 
considering also that other more standard methods suffer not only for poor accuracy but also for the too large extension of the  
basis set. It also worth noticing that  we obtain an higher gain in the region $r_x \approx r_y$ where the RVB picture is more relevant.

\begin{figure}[t]
    \includegraphics[scale=0.8]{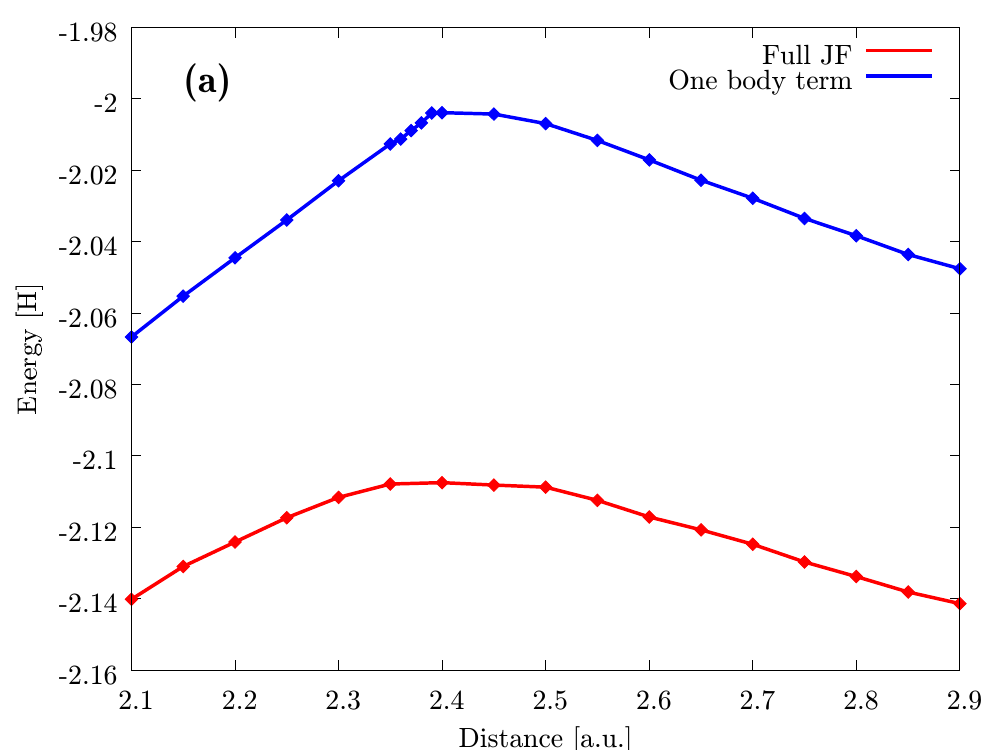}
    \includegraphics[scale=0.8]{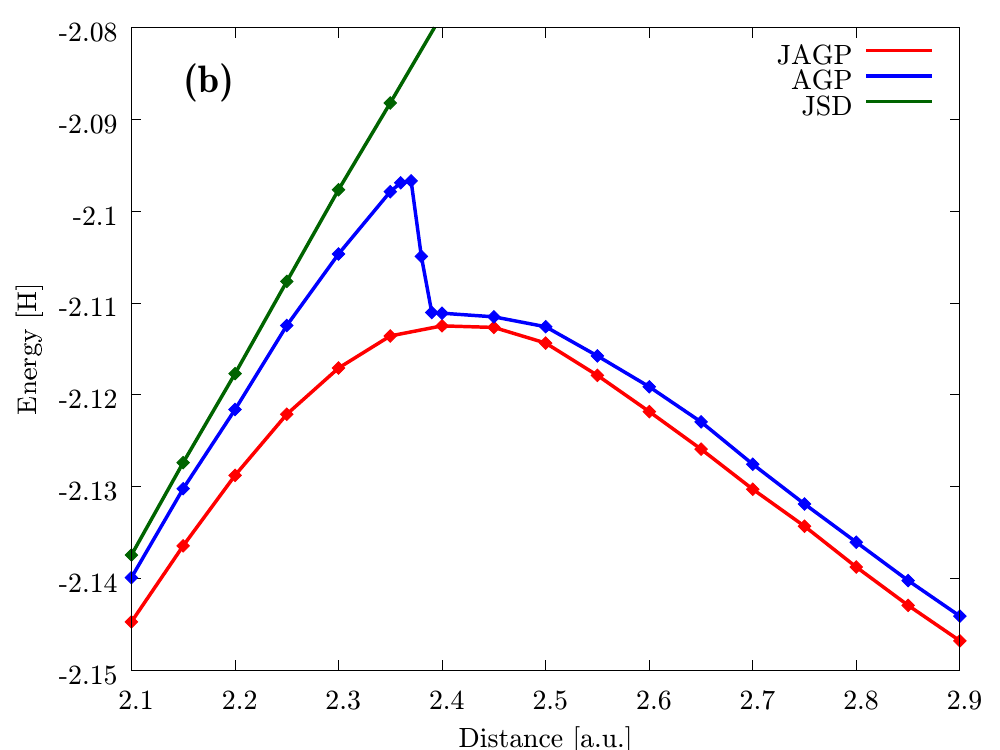}
    \caption{AGP and JAGP energies. In panel (\textbf{a}) we compare the values at the
      VMC level, while in  panel \textbf{b}  the corresponding DMC energies, within FNA,  are shown.}
    \label{jasnojas}
\end{figure}

Thanks to the simplicity of the $H_4$ molecule, and the limited number of variational 
parameters involved in our  WFs, we can use this model to study the genuine AGP without 
any JF. This case is particularly difficult with 
our stochastic optimization method because the statistical fluctuations of the energy are much larger.
In principle the AGP should be
able to describe the static correlation  of this molecule also without JF, with the two main 
contributions in the WF 
shown pictorially in Fig.~\ref{denspl}\textbf{a}. At the variational level a much worse energy for the AGP
WF is expected  bacause the correlation 
described by the JF is very important. However, it is very interesting to observe that 
the DMC results are  significantly 
different with (JAGP) or without (AGP) JF, even considering that 
the  JF$>0$ cannot change  the signs of the WF, and only the optimization of the AGP  in presence of the JF leads to 
a very accurate nodal surface.
In Fig.~\ref{jasnojas}\textbf {a}  we can see that the variational energies of the AGP WF are indeed considerably higher compared 
to the JAGP ones. The smoothness of the curve and the reproducibility of the results
indicate that the optimization is not stuck in spurious local minima.
Instead  the DMC results shown in  Fig.~\ref{jasnojas}\textbf {b} indicate  
an unphysical jump of the energy between two different phases. 
When $r_x \ge r_y$ the AGP is able
to give very good  energies that differ only few mH from the JAGP ones. Instead, when $r_x < r_y$ we can see a clear jump
in the energy indicating that the nodal surface of the WF is not correctly described by the AGP. However
also in this regime the nodes are still better that the ones provided by the JSD WF, with energy values between the ones of the
JSD and the JAGP. In order to check that this transition was not due to some optimization error we have calculated the WFs for
$r_x<r_y$ starting from $r_x=r_y$, obtaining exactly the same VMC and DMC results. Qualitatively speaking, when the AGP is optimized
in presence of the JF, it can resonate between the correct configurations by avoiding double occupancies of singlet electron pairs
\cite{rvb_pauling,rvb_pwa}, that   
 are  energetically unfavorable. In some sense the Jastrow correlation drives the optimization of the AGP toward 
the correct ground state energy and nodal surface.

Finally, as we can see from Table (\ref{tab:varen}), the JAGP is almost converged to the CBS limit 
with only the double zeta cc-pVDZ basis. The differences in energy with the cc-pVTZ are much below one mH per atom. 
This fast convergence is  due to the term in the Eq.~(\ref{sb1}) that fulfills the electron-ion cusp conditions and 
allows us to use a very small basis set to describe the system. In the AGP the  number of variational parameters scales with 
the square of the number of elements of the basis. It is therefore very important to reach a very accurate description 
with the smallest possible basis set. This can have a very dramatic impact for large systems where the dimension of the basis 
set is one of the most important bottlenecks of our JAGP calculations.

\section{Conclusion}
In this work we have applied state of  the art QMC techniques to  a very simple system that has been used for benchmarking
their accuracy in describing the strong electron correlations. We have shown that the 
use of the JAGP wave function is not only qualitatively correct but allows an almost exact description of the ground state 
with a computational 
effort similar to the widely used JSD, that miserably fails in this system, even  within the
more accurate DMC approach. We have also
shown that the full optimization of our JAGP ansatz guarantees a very fast convergence in the basis set, so that no kind of  
extrapolation is necessary for almost converged  results in the CBS limit. Considering the above remarkable properties of our JAGP ansatz, 
the extension to larger systems has been already employed in several cases, but its accuracy has not been deeply 
investigated. 
Thanks to the simplicity of this  model system we were also able to prove that the AGP alone, without the use of a JF, 
miserably fails,  leading not only to inaccurate DMC energies but qualitatively wrong, as  a discontinuity of 
the energy landscape as a function of the atomic positions was reported. In this case the wrong  nodal
 surface determined  by the AGP for small $r_x$,  was not detectable at the VMC level,  
 because the optimized VMC energy was indeed a smooth and continuous function of $r_x$, as it should be from general grounds.
 This example leads us to conclude that the AGP optimization, in principle possible 
 with a very cheap and deterministic algorithm similar to the HF self consistent method, is 
 completely useless, as it can lead to  spurious and qualitatively wrong results, in this sense much worse than simpler HF or DFT calculations.
This work clearly indicates that instead the JAGP ansatz opens the way to tackle even more complicated systems, when standard 
quantum chemistry methods are too much expensive and the single determinant approach does not work well. 

\bibliography{Bibliography}


\end{document}